\documentstyle[aaai]{article}

\newcommand{\fix}[1]{#1\!\uparrow}

\newcommand{\Si}{\Sigma}

\newcommand{\n}{\mbox{\bf not}}
\newcommand{\Tr}{\mbox{\bf t}}
\newcommand{\Fa}{\mbox{\bf f}}
\newcommand{\Un}{\mbox{\bf u}}

\newcommand{\HH}[1]{{\cal H}_{#1}}
\newcommand{\lep}{\leq_p}

\newcommand{\leI}{\leq_{kn}}

\newcommand{\geT}{\geq_{tr}}

\newcommand{\All}{{\cal A}}
\newcommand{\Ball}{{\cal B}}

\newcommand{\Least}{\bot}

\newcommand{\DER}[1]{{\cal D}_{#1}}

\newcommand{\np}{\overline{p}}

\newcommand{\proves}{\vdash}

\newcommand{\Proj}[1]{\mbox{\it Proj}({#1})}

\newcommand{\inst}[2]{#2_{#1}}

\newcommand{\Belp}[1]{{\cal B}el(#1)}
\newcommand{\oath}[1]{(#1)^{\mathit wk}}
\newcommand{\uath}[1]{(#1)^{\mathit str}}

\newcommand{\SDER}[1]{{\cal SD}_{#1}}
\newcommand{\aela}[1]{ael_1({#1})}
\newcommand{\aelb}[1]{ael_2({#1})}

\newenvironment{example}{\begin{nexample}}{\end{nexample}}

\newtheorem{definition}{Definition}
\newtheorem{proposition}{Proposition}
\newtheorem{theorem}{Theorem}

\newtheorem{corollary}{Corollary}

\newtheorem{nexample}{Example}

\begin{document}

\title{Fixpoint 3-valued semantics for autoepistemic logic}

\author{
{\it Marc Denecker}\\
Department of Computer Science\\
K.U.Leuven\\
Celestijnenlaan 200A, B-3001 Heverlee, Belgium\\
{\tt marcd@cs.kuleuven.ac.be}
\And
{\it Victor Marek and Miros\l aw Truszczy\'nski}\\
Computer Science Department\\
University of Kentucky\\
Lexington, KY 40506-0046\\
{\tt marek|mirek@cs.engr.uky.edu}
}

\maketitle

\begin{abstract}
\begin{quote}
The paper presents a constructive 3-valued semantics for autoepistemic 
logic (AEL). We introduce a derivation operator and define the
semantics as its least fixpoint. The semantics is 3-valued in 
the sense that, for some formulas, the least fixpoint does not specify 
whether they are believed or not. We show that complete fixpoints 
of the derivation operator correspond to Moore's stable expansions. 
In the case of modal representations 
of logic programs our least fixpoint semantics expresses well-founded 
semantics or 3-valued Fitting-Kunen semantics (depending on 
the embedding used). We show that, computationally, our semantics is 
simpler than the semantics proposed by Moore (assuming that 
the polynomial hierarchy does not collapse).
\end{quote}
\end{abstract}

\vspace*{-2mm}
\section{Introduction}
\label{intro}

We describe a 3-valued semantics for modal theories that
\renewcommand{\thefootnote}{\fnsymbol{footnote}}
\footnote[0]{\hspace*{-6mm}Copyright
\copyright 1998, American Association for Artificial Intelligence
(www.aaai.org). All rights reserved.}
\renewcommand{\thefootnote}{\arabic{footnote}}
approximates skeptical mode of reasoning in the autoepistemic logic
introduced in \cite{mo84,mo85}. We present results 
demonstrating that our approach is, indeed, appropriate for modeling 
autoepistemic reasoning. We discuss computational properties of 
our semantics and connections to logic programming.

Autoepistemic logic is among the most extensively studied nonmonotonic 
formal systems. It is closely related to default logic \cite{re80}.
It can handle default reasonings under a simple and modular translation in 
the case of prerequisite-free defaults \cite{mt93}. In the case of 
arbitrary default theories, a somewhat more complex non-modular 
translation provides a one-to-one correspondence between default 
extensions and stable (autoepistemic) expansions \cite{got95}. Further, 
under the so called Gelfond translation, autoepistemic logic captures 
the semantics of stable models for logic programs \cite{ge87}. Under 
the Konolige encoding of logic programs as modal theories, stable 
expansions generalize the concept of the supported model 
semantics \cite{mt93}. Autoepistemic logic is also known to 
be equivalent to several other modal nonmonotonic reasoning
systems.

The semantics for autoepistemic logic \cite{mo85} 
assigns to a modal theory $T$ a collection of its {\em stable 
expansions}. This collection may be empty, may consist of exactly one
expansion or may consist of several different expansions.
Intuitively, expansions are designed to model 
belief states of agents with {\em perfect} introspection powers: 
for every formula $F$, either the formula $KF$ (expressing 
a belief in $F$) or the formula $\neg KF$ (expressing that
$F$ is not believed) belongs to an expansion. We will say that
expansions contain no {\em meta-ignorance}. 

In many applications, the phenomenon of multiple expansions is desirable.
There are situations where we are not interested in answers
to queries concerning a single atom or formula, but in a {\em
collection} of atoms or formulas that satisfy some constraints. Planning
and diagnosis in artificial intelligence, and a range of 
combinatorial optimization problems, such as computing hamilton
cycles or $k$-colorings in graphs, are of this type. These problems may
be solved by means of autoepistemic logic precisely due to the fact that
multiple expansions are possible. The idea is to represent a problem as
an autoepistemic theory so that solutions to the problem are in
one-to-one correspondence with stable expansions. While
conceptually elegant, this approach has its problems. Determining
whether expansions exist is a $\Sigma^P_2$-complete problem 
\cite{nie92,got92}, and all known algorithms for computing expansions 
are highly inefficient.

In a more standard setting of knowledge representation, the goal is to 
model the knowledge about a domain as a theory in some formal system 
and, then, to use some inference mechanism to resolve queries against 
the theory or, in other words, establish whether particular formulas are 
entailed by this theory. Autoepistemic logic (as well as other 
nonmonotonic systems) can be used in this mode, too. Namely, under 
the so called {\em skeptical} model, a formula is entailed by 
a modal theory, if it belongs to all stable expansions of this theory. 
The problem is, again, with the computational complexity of determining 
whether a formula belongs to all expansions. 

We propose an alternative semantics for autoepistemic
reasoning that allows us to {\em approximate} the skeptical 
approach described above. Our semantics has the property that 
if it assigns to a formula the truth value $\Tr$, then this formula 
belongs to all stable expansions and, dually, if it assigns to a formula 
the truth value $\Fa$, then this formula does not belong to any
expansion. Our semantics is 3-valued 
and some formulas are assigned the truth value $\Un$ (unknown).  While 
only approximating the skeptical mode of reasoning, it has one important 
advantage. Its computational complexity is lower (assuming that 
the polynomial hierarchy does not collapse on some low level). Namely, 
the problem to determine the truth value of a formula is in 
the class $\Delta^P_2$.

Clearly, the semantics we propose can be applied whenever the situation
requires that autoepistemic logic be used in the skeptical 
mode. However, it has also another important application. It can be used
as a pruning mechanism in algorithms that compute expansions. The idea
is to first compute our 3-valued interpretation for an autoepistemic
theory (which is computationally simpler than the task of computing an
expansion) and, in this way, find some formulas which are in all
expansions and some that are in none. This restricts the search space
for expansions and may yield significant speedups.

Conceptually, our semantics plays the role similar to that played by the
well-founded semantics in logic programming. Deciding whether an
atom is in all stable models is a co-NP-complete problem. However, the
well-founded semantics, which approximates the stable model semantics
can be computed in polynomial time. Furthermore, well-founded semantics 
is used both as the basis for top-down query answering implementations of 
logic programming \cite{csw94}, and as a search space pruning 
mechanism by some implementations to compute stable model semantics 
\cite{ns95}.

Our 3-valued semantics for autoepistemic logic is based on the
notion of a {\em belief pair}. These are pairs $( P, S)$, where $P$ 
and $S$ are sets of 2-valued interpretations of the underlying 
first-order language, and $S \subseteq P$. 
The motivation to consider belief pairs comes from Moore's 
possible-world characterization of stable expansions \cite{mo84}. 
Moore characterized consistent expansions in terms of {\em possible-world
structures}, that is, non-empty sets of 2-valued interpretations.
A belief pair $(P,S)$ can be viewed as an approximation to a
possible-world structure $W$ such that $S\subseteq W\subseteq P$: 
interpretations not in $P$ are known not to be in $W$, and those in 
$S$ are known to be in $W$. Observe that while expansions (or 
the corresponding possible-world structures) do not contain meta-ignorance, 
belief pairs, in general, do. 

Our semantics is defined in terms of fixpoints of a 
monotone operator defined on the set of belief pairs. 
This operator, $\DER{T}$, is determined by an initial theory $T$.
Given a belief pair $B=(P,S)$, $\DER{T}$ establishes that some
interpretations that are in $P$ must, in fact, belong to $S$. 
In addition, some other interpretations 
in $P$ are eliminated altogether, as inappropriate for describing
a possible state of the world (given the agent's initial knowledge). 
The operator attempts to simulate a constructive process rational agents 
might use to produce an ``elementary'' improvement on their current set 
of beliefs and disbeliefs. 

We say that $( P_1,S_1)$
``better approximates'' the agent's beliefs and disbeliefs entailed by
the agent's initial assumptions than $(P,S)$ if
$S\subseteq S_1\subseteq P_1\subseteq P.$
The operator $\DER{T}$ is monotone with respect to this
ordering and, thus, it has the least fixpoint. 
We propose this fixpoint as a {\em constructive} 
approximation to the semantics of stable expansions.

A fundamental property that makes the above construction meaningful is
that {\em complete} belief pairs (those with $P$ equal to $S$) that 
are fixpoints of $\DER{}$ are precisely Moore's S5-models characterizing 
expansions. Thus, by the general properties of fixpoints of monotone
operators over partially ordered sets, the least fixpoint
described above indeed approximates the skeptical reasoning based on
expansions. Moreover, as mentioned above, the problem of computing
the least fixpoint of the operator $D$ requires only polynomially many 
calls to the satisfiability testing engine, that is, it is in $\Delta_2^P$.
Another property substantiating our approach is that under
some natural encodings of logic programs as modal theories, our
semantics yields both well-founded semantics \cite{vrs91}
and the 3-valued Fitting-Kunen semantics \cite{fi85,kun87}.


\vspace*{-2mm}
\section{Autoepistemic logic --- preliminaries}
 \label{SecPreliminaries}

The language of autoepistemic logic is the standard language of
propositional modal logic over a set of atoms $\Sigma$ and with a single
modal operator $K$. We will refer to this language as ${\cal L}_K$. The
propositional fragment of ${\cal L}_K$ will be denoted by $\cal L$.

The notion of a 2-{\em valued interpretation}
of the alphabet $\Sigma$ is defined as 
usual: it is a mapping from $\Sigma$ to $\{\Tr,\Fa\}$. 
The set of all interpretations of $\Si$ is denoted 
$\All_{\Si}$ (or $\All$, if $\Si$ is clear from the context).

%

A possible-world semantics for autoepistemic logic was introduced by
Moore \cite{mo84} and proven equivalent with the semantics of stable 
expansions.
A possible-world structure $W$ (over $\Sigma$) is a set of
2-valued interpretations of $\Sigma$. Alternatively, it can be seen 
as a Kripke structure with a total accessibility relation.
Given a pair $(W,I)$, where $W$ is a set of
interpretations and $I$ is an interpretation (not necessarily from $W$), 
one defines a truth 
assignment function $\HH{W,I}$ inductively as follows:\\
(1) For an atom $A$, we define $\HH{W,I}(A) = I(A)$;\\ 
(2) The boolean connectives are handled in the usual way;\\ 
(3) For every formula $F$, we define $\HH{W,I}(KF) = \Tr$ if 
for every $J\in W, \HH{W,J}(F)=\Tr$, and $\HH{W,I}(KF) = \Fa$,
otherwise.

We write $(W,I) \models F$ to denote that $\HH{W,I}(F)=\Tr$.
Further, for a modal theory $T$, we will write $(W,I) \models T$ if
$\HH{W,I}(F)=\Tr$ for any $F \in T$. Finally, for a possible 
world structure $W$ we define the {\em theory} of $W$, $Th(W)$, by:
$Th(W)=\{F\colon (W,I)\models F,\ \mbox{for all $I\in W$}\}$.

It is well known that for every formula $F$, either $KF\in Th(W)$
or $\neg KF\in Th(W)$ ($\HH{W,I}(KF)$ is the same for all 
interpretations $I \in W$). Thus, possible-world structures have no 
meta-ignorance and, as such, are suitable for modeling belief sets of 
agents with {\em perfect} introspection capabilities. It is precisely this 
property that made possible-world structures fundamental objects in the
study of modal nonmonotonic logics \cite{mo84,mt93}. 

\begin{definition}
An {\em autoepistemic model} of a modal
theory $T$ is a possible-world ({\bf S5}) structure $W$ which satisfies 
the following fixpoint equation:
\vspace*{-1mm}
\[
W = \{I \colon (W,I)\models T\}.
\]
\end{definition}

\vspace*{-1mm}
The following theorem, relating stable expansions of \cite{mo85}
and autoepistemic models, was proved in \cite{le90} and 
was discussed in detail in \cite{sch91b}.

\begin{theorem}\label{PStableExpAutoModel} 
For any two modal theories $T$ and $E$, $E$ is a consistent stable 
expansion of $T$ if and only if $E=Th(W)$ for some nonempty 
autoepistemic model $W$ of $T$.
\end{theorem}

\vspace*{-2mm}
\section{A fixpoint 3-valued semantics for autoepistemic logic}
\label{SecDER}

Our semantics for autoepistemic logic is defined in terms of
possible-world structures and fixpoint conditions. The key difference 
with the semantics proposed by Moore is that we consider 
{\em approximations} of possible-world structures by 
{\em pairs} of possible-world structures. Recall from the previous 
section, that $\All$ denotes the set of all interpretations of a fixed 
propositional language $\cal L$.

\begin{definition}
\label{Dbeliefpair}
A {\em belief pair} is a pair $(P,S)$ of sets of interpretations
$P$ and $S$ such that $P \supseteq S$. When
$B=(P,S)$, $S(B)$ denotes $S$ and $P(B)$ denotes $P$.
The belief pair $(\All,\emptyset)$ is denoted $\Least$. The set
$\{ (P,S)\colon P, S \in \All \mbox{\ and\ } P \supseteq S\}$ of all
belief pairs is denoted by $\Ball$. 
\end{definition}

The interpretations in $S(B)$ can be viewed as states of
the world which are known to be possible (belong to $W$).
They form a lower approximation to $W$. The interpretations in $P(B)$ 
can be viewed as an upper approximation to $W$. In other words,
interpretations not in $P(B)$ are known not to be in $W$. 

We will extend now the concept of an interpretation to the
case of belief pairs and consider the question of meta-ignorance and
meta-knowledge of belief pairs. We will show that, being only
approximations to possible-world structures, belief pairs may contain
meta-ignorance. We will use three logical values, $\Fa$, $\Un$ 
and $\Tr$. In the definition, we will use the {\em truth} 
ordering: $\Fa \leq_{tr} \Un \leq_{tr} \Tr$ and define 
$\Fa^{-1}=\Tr, \Tr^{-1}=\Fa, \Un^{-1}=\Un$.

\begin{definition}
 \label{Dtruthfunction}
 Let $B = (P,S)$ be a belief pair and let $I$ be an interpretation. The 
truth function $\HH{B,I}$ is defined inductively:

 \begin{tabbing}\hspace{5mm} \=\+ $\HH{I}(p(t_1,\ldots,t_n)$ \= \kill
$\HH{B,I}(A)$   \>      $= I(A)$ \ \ (A is an atom) \\
$\HH{B,I}(\neg F)$      \>      $= \HH{B,I}(F)^{-1}$ \\
$\HH{B,I}(F_1 \lor F_2)$ \>     $= max\{\HH{B,I}(F_1),
\HH{B,I}(F_2)\}$\\
$\HH{B,I}(F_1 \land F_2)$\>     $= min\{\HH{B,I}(F_1),
\HH{B,I}(F_2)\}$\\
$\HH{B,I}(F_2 \supset F_1)$\>   $= max\{\HH{B,I}(F_1),
\HH{B,I}(F_2)^{-1}\}$
 \end{tabbing}
The formula $K(F)$ is evaluated as follows:
\vspace*{-2mm}
\[
\HH{B,I}(K(F))= \left \{
\begin{array}{ll}
 \Tr \ \ \mbox{if $\forall_{J \in P} \HH{B,J}(F)=\Tr$}\\
 \Fa \ \ \mbox{if $\exists_{J \in S} \HH{B,J}(F)=\Fa$}\\
 \Un \ \ \mbox{otherwise}
\end{array}
\right.
\]
\end{definition}

The truth value of a modal atom
$KF$, $\HH{B,I}(KF)$, does not depend on the choice of
$I$. Consequently, for
a modal atom $KF$ we will write $\HH{B}(KF)$ to denote this,
common to all interpretations from $\All$, truth value of $KF$.

Let us define the {\em meta-knowledge} of a belief
pair $B$ as the set of formulas $F\in {\cal L}_K$ such that 
$\HH{B}(KF) =\Tr$ or $\HH{B}(KF) =\Fa$. The {\em meta-ignorance}
is formed by all other formulas, that is, those formulas $F\in {\cal
L}_K$ for which $\HH{B}(KF) =\Un$.


Clearly, a belief pair $B=(W,W)$ naturally corresponds to a
possible-world structure $W$. Such a belief pair is called {\em
complete}. We will denote it by $(W)$. The following straightforward 
result indicates that $\HH{B,I}$ is a generalization of $\HH{W,I}$ 
to the case of belief pairs. It also states that a complete belief 
pair contains no meta-ignorance.

\begin{proposition}
\label{PTF2valued}
If $B$ is a complete belief pair $(W)$, then $\HH{B,I}$ is 2-valued.
Moreover, for every formula $F$, $\HH{B,I}(F) = \HH{W,I}(F)$.
\end{proposition}

In our approach to autoepistemic reasoning we will model the agent who, 
given an initial theory $T$, starts with the belief pair $\Least$
(with the smallest meta-knowledge content) and, then, iteratively
constructs a sequence of belief pairs with increasing meta-knowledge
(decreasing meta-ignorance) until no more improvement is possible.
To this end, we will introduce now a partial ordering on the set $\Ball$ 
of belief pairs. Given two belief pairs $B_1$ and $B_2$, we define 
$B_1 \lep B_2$ if $P(B_1) \supseteq P(B_2)$ and $S(B_1) \subseteq S(B_2)$. 
This ordering is consistent with the ordering defined by the ``amount'' of 
meta-knowledge contained in a belief pair: the "higher" a belief pair
in the ordering $\lep$, the more meta-knowledge it contains (and the 
less meta-ignorance). It is also consistent with the concept of the 
{\em information ordering} of the truth values: $\Un \leI
\Fa, \Un \leI \Tr$, $\Fa\not\leI\Tr$ and $\Tr\not\leI\Fa$. 

\begin{proposition}
\label{mt-1}
Let $B_1$ and $B_2$ be belief pairs. If $B_1\lep B_2$ then
for every $F\in{\cal L}_K$ and for every interpretation $I$, 
$\HH{B_1,I}(F) \leI \HH{B_2,I}(F)$. 
\end{proposition}

The ordered set $(\Ball,\lep )$ is not a lattice. In fact, for every
$W \subseteq {\cal A}$, $(W)$ is a maximal element in $(\Ball,\lep )$. 
If $W_1 \neq W_2 $, then $(W_1)$ and $(W_2)$ have no least upper bound
(l.u.b.) in $(\Ball,\lep )$. The pair $\bot = ({\cal A}, \emptyset )$ is 
the least element of $(\Ball,\lep )$.

The ordered set $(\Ball,\lep )$ is {\em chain-complete}. That is, every
set of pairwise comparable elements has the l.u.b. A monotone operator 
defined on a chain-complete ordered set with a least element has 
a least fixpoint, which is the limit of the iterations of the operator 
starting at the least element $\bot$ \cite{mrk76}. 

We will now define a monotone operator on the ordered set
$(\Ball,\lep)$ and will use it to define a step-wise process of 
constructing belief pairs with increasing meta-knowledge. To this end,
we will define two satisfaction relations: {\em weak} (denoted by 
$\models_w$) and {\em strong} (denoted by $\models$):
\begin{tabbing} \hspace{5mm}\=\+
$(B,I) \models_w F$ if $\HH{B,I}(F) \not=\Fa$ (i.e. $\HH{B,I}(F)\geT \Un$)\\
$(B,I) \models F$ if $\HH{B,I}(F) = \Tr$\\
\end{tabbing}

\vspace*{-4mm}
\begin{definition}
 \label{DERerivation}
Given $B \in \Ball$, the value of the derivation operator $\DER{T}$ is
defined as follows:
\vspace*{-2mm}
\[
\DER{T}(B) = (\{I\ |\ B,I \models_w T\}, \{I\ |\ B,I \models T\}). 
\]
\end{definition}

\vspace*{-1mm}
Thus, $P(\DER{T}(B))$ consists of the states which weakly satisfy 
$T$, according to $B$, while $S(\DER{T}(B))$ are the states which strongly 
satisfy $T$ according to $B$. The subscript $T$ is omitted when $T$
is clear from the context.

\begin{example} \label{ExT0.1}
{\rm
Consider $T = \{ K(p) \supset q \}$. 
Then $\DER{}(\Least) = (\All, \{ pq, \np q\})$. Indeed,
$\HH{\Least}(Kp) = \Un$. Consequently, for every $I$, 
$\HH{{\Least},I}(Kp\supset q) \not=\Fa$, that is,
$(\Least,I)\models_w Kp\supset q$. For the same reason,
$\HH{{\Least},I}(Kp\supset q)) = \Tr$ if and only if $I(q)=\Tr$. 
To compute $\DER{}^2(\Least)$, observe that 
$\HH{\DER{}(\Least)}(Kp) =\Fa$. Consequently, for every $I$,
$\HH{\DER{}(\Least),I}(Kp\supset q) = \Tr$. It follows that
$\DER{}^2(\Least) = (\All,\All)$. It is also easy to see now that
$(\All,\All)$ is the fixpoint of $\DER{}$. Notice that the belief pair
$(\All)$, obtained by iterating $\DER{}$, corresponds to the
possible-world structure $\All$ that defines the unique stable expansion
of $T$.
}
 \end{example}

Basic properties of the operator $\DER{}$ are gathered in the following
proposition.

\begin{proposition}
\label{PDERmonotonic} 
Let $T$ be a propositionally consistent modal theory. Then, for every belief 
pair $B$:\\
(1) $\DER{T}(B)$ is a belief pair.\\
(2) $\DER{T}$ is monotone on $\Ball$.\\
(3) If $B$ is complete, then $\DER{T}(B)$ is complete.
\end{proposition}

Since $(\Ball,\lep)$ is a chain-complete ordered set with least element
$\Least$ and $\DER{}$ is monotonic, $\DER{}$ has a least 
fixpoint. We will denote it by $\DER{}\uparrow$. We propose 
this fixpoint as the semantics of modal theories. This semantics reflects 
the reasoning process of an agent who gradually constructs belief pairs
with increasing information content. In the remainder of the paper, we
will study properties of this semantics and, more generally, of fixpoints 
of the operator $\DER{}$. The next three results relate fixpoints of 
$\DER{}$ to Moore's semantics.

\begin{theorem} \label{PAEMDER}
Let $T\subseteq {\cal L}_K$. A possible-world structure $W$ is 
an autoepistemic model of $T$ if and only if $(W)$ is a 
fixpoint of $\DER{T}$. If ${\DER{T}}\uparrow$ is complete then it is the
unique autoepistemic model of $T$.
 \end{theorem}

Using Propositions \ref{PStableExpAutoModel} and \ref{mt-1}, we can
extract a relationship between stable expansions and fixpoints of the
derivation operator $\DER{T}$.

\begin{corollary} For any pair $T, E$ of modal theories, 
$E$ is a consistent stable expansion of $T$ if and only if $E=Th(S)$ for 
some complete fixpoint $(S)$ of $\DER{T}$.\\
If $\HH{{\cal D}_T\uparrow} (F) = \Tr$ then $F$ 
belongs to all expansions of $T$. If $\HH{{\cal D}_T\uparrow} (F) = \Fa$
then $F$ does not belong to any expansion of $T$.
\end{corollary}

Consistent stratified autoepistemic theories \cite{ge87} have a unique
autoepistemic model (stable expansion). Our semantics coincides with the
Moore's semantics on stratified theories.
 \begin{theorem} \label{ThStatifiedAEL1} If $T$ is a consistent
stratified autoepistemic theory, then $\fix{\DER{}}$ is complete. Hence,
it is the unique autoepistemic model of $T$.
 \end{theorem}

Thus, the semantics defined by the least fixpoint of the operator $\DER{}$
has several attractive features. It is defined for every consistent
modal theory $T$. It coincides with the semantics of autoepistemic
logic on stratified theories and, in the general case, provides 
an approximation to all stable expansions (or, in other words, to
skeptical autoepistemic reasoning).

%
%

\vspace*{-2mm}
\section{An effective implementation of $\DER{}$} \label{imp}

The approach proposed and discussed in the previous section does not
directly yield itself to fast implementations. The definition of the
operator $\DER{}$ refers to all interpretations of the language $\cal
L$. Thus, computing $\DER{}(B)$ by following the definition is
exponential, even for modal theories of a very simple syntactic form. 
Moreover, representing belief pairs is costly. Each of the sets
$P(B)$ and $S(B)$ may contain exponentially many elements.

In this section, we describe a characterization of the operator
$\DER{}$ that is much more suitable for investigations of algorithmic
issues associated with our semantics. The strategy is to represent a 
belief set $B$ as a theory $Rep(B)$. Since the theory $Rep (B)$ needs 
to represent {\em two} sets of valuations, $Rep (B)$ will be a theory 
in the propositional language extended by three constants $\Tr$, $\Fa$
and $\Un$. These constants will always be interpreted by the logical 
values $\Tr$, $\Fa$ and $\Un$, respectively. We will call such theories 
3-FOL theories.

Let $F$ be a 3-FOL formula. By $\oath{F}$ we denote the formula obtained 
by substituting $\Tr$ for all positive occurrences of $\Un$ and $\Fa$ 
for all negative occurrences of $\Un$. Similarly, by $\uath{F}$ we denote 
the formula obtained by substituting $\Tr$ for all negative occurrences of 
$\Un$ and $\Fa$ for all positive occurrences of $\Un$. Given a 3-FOL theory
$Y$, we define $\uath{Y}$ and $\oath{Y}$ by standard setwise extension.

Clearly, $\uath{F}$ and $\oath{F}$ do not contain $\Un$.
Consequently, they can be regarded as formulas in the propositional 
language extended by two constants $\Tr$ and $\Fa$ with standard 
interpretations as truth and falsity, respectively. We will call this 
language 2-FOL. We will write $\proves$ and $\models$ to denote 
provability and entailment relations in 2-FOL. An important observation
here is that if an interpretation satisfies the formula $\uath{F}$ then 
it also satisfies $\oath{F}$. That is $\uath{F}\supset\oath{F}$ is a
tautology of 2-FOL.  For a 2-FOL theory $U$, we define:
$Mod(U) = \{ I : \ \mbox{ for all $F \in U,\ I\models F$}\}.$
It follows that for a 3-FOL theory $Y$, $Mod(\uath{Y}) \subseteq 
Mod(\oath{Y})$. Thus, $(Mod(\oath{Y}),Mod(\uath{Y})$ is a belief pair 
and $Y$ can be viewed as its representation. 

We show now how, similarly to belief pairs, 3-FOL theories can be used 
to assign truth 
values to {\em modal} atoms (and, hence, to all modal formulas). Let $Y$ 
be a 3-FOL theory, and let $F$ be a modal formula. Define $\HH{Y}(K(F))$
by induction of depth of formula $F$ as follows:\\
(1) If $F$ is objective, then define:
\vspace*{-2mm}
\[
\HH{Y}(K(F))= \left \{
\begin{array}{ll}
 \Tr \ \ \mbox{if $\oath{Y} \proves \uath{F}$}\\
 \Fa \ \ \mbox{if $\uath{Y} \not \proves \oath{F}$}\\
 \Un \ \ \mbox{otherwise.}
\end{array}
\right.
\]
(2) If $F$ is not objective, then replace all modal atoms $K(G)$
in $F$ by $\HH{Y}(K(G))$. This yields an objective formula $F'$. 
Define $\HH{Y}(K(F)) =\HH{Y}(K(F'))$.\\

Let $T$ be a modal theory and let $Y$ be a 3-FOL theory. By the
$Y$-instance of $T$, $T_Y$, we mean a 3-FOL theory obtained by
substituting all modal literals $K(F)$ (not appearing under the scope
of any other occurrence of $K$) by $\HH{Y}(K(F))$. 
Observe that for a finite modal
theory $T$ and a finite 3-FOL theory $Y$, $\inst{Y}{T}$ can be computed
by means of polynomially many calls to the propositional provability
procedure. 

Let $T$ be a modal theory. We will now define a counterpart to 
the operator $\DER{T}$. Let $Y$ be a 3-FOL theory. Define 
$\SDER{T}(Y) = \inst{Y}{T}.$


The key property of the operator $\SDER{T}$ is that, for a finite modal
theory $T$ and for a finite 3-FOL theory $Y$, $\SDER{T}(Y)$ can be
computed by means of polynomially many calls to the propositional
provability procedure.

We will show that $\SDER{T}$ can be
used to compute $\DER{T}$. In particular, we will show that the least
fixpoint of $\DER{T}$ can be computed by iterating the operator
$\SDER{T}$. To this end, for every 3-FOL theory $Y$, define $\Belp{Y} = 
(Mod{(\oath{T})},Mod{(\uath{T})})$.

First, the following theorem shows that the truth values of modal atoms 
evaluated according to a 3-FOL theory $T$ and according to the
corresponding belief pair $Bel(T)$ coincide.
 
\begin{theorem} \label{THHBHHT} 
Let $Y$ be a 3-FOL theory. Then, for every modal formula $F$,
\vspace*{-2mm}
\[
\HH{Bel(Y)}(K(F)) = \HH{Y}(K(F)).
\]
\end{theorem}

Next, let us observe that the operator $\DER{}$ can be described in
terms of the operator $Bel$. Let $T$ be a modal theory and let $B$
be a belief pair. By the $B$-instance of $T$, $T_B$, we mean a 3-FOL theory 
obtained by substituting all modal literals $K(F)$ (not appearing under 
the scope of any other occurrence of $K$) by $\HH{B}(K(F))$. 

\begin{theorem} \label{TDERInst}
Let $T$ be a modal theory and let $B$ be a belief pair. Then,
$\DER{T}(B) = \Belp{\inst{B}{T}}$
\end{theorem}

This theorem indicates that, given a modal theory $T$, belief pairs that
are in the range of the operator $\DER{T}$ can be represented by objects
of size polynomial in the size of $T$. Namely, every belief pair of the
form $\DER{T}(B)$ can be represented by a 3-FOL theory $T_B$. 

Theorems \ref{THHBHHT} and \ref{TDERInst} imply the main result of this
section.

\begin{theorem} \label{TDERSDER} 
Let $T$ be a modal theory and let $Y$ be a 3-FOL theory.\\
(1) $\Belp{\SDER{T}(Y)} = \DER{T}(Bel(Y))$.\\
(2) If a belief pair $B$ is a fixpoint of $\DER{T}$, then 
$\inst{B}{T}$ is a fixpoint of $\SDER{T}$.\\  
(3) If $Y$ is a fixpoint of $\SDER{T}$ then $\Belp{Y}$ is a fixpoint
of $\DER{T}$.
\end{theorem}

Observe that $Bel(\{\Un\}) = \Least$. It follows directly from 
Theorem \ref{TDERSDER} (by induction) that for every ordinal
number $\alpha$, $\DER{T}^{\alpha}(\Least) = 
Bel(\SDER{T}^{\alpha}(\{\Un\})).$

Clearly, if $\SDER{T}^{\alpha}(\Least) = \SDER{T}^{\alpha+1}(\Least)$ 
then $\DER{T}^{\alpha}(\Least) = \DER{T}^{\alpha+1}(\Least)$. Moreover,
by Theorems \ref{THHBHHT}, \ref{TDERInst} and \ref{TDERSDER}, and 
by induction, it is easy 
to show that if $\DER{T}^{\alpha}(\Least) = \DER{T}^{\alpha+1}(\Least)$ 
then $\SDER{T}^{\alpha+1}(\Least) = \SDER{T}^{\alpha+2}(\Least)$.

Consequently, the least fixpoint of $\DER{T}$ (its polynomial-size
representation) can be computed 
by iterating the operator $\SDER{T}$. In the case when $T$ is finite,
the number of iterations is limited by the number of top level
(unnested) modal literals in $T$. Originally, they may all be evaluated
to $\Un$. However, at each step, at least one $\Un$ changes to either 
$\Tr$ or $\Fa$ and this value is preserved in the subsequent
evaluations. Thus, the problem of computing a polynomial size 
representation of the least fixpoint of the operator $\DER{}$, 
the corresponding 3-FOL theory, is in the class $\Delta^P_2$.

\section{Relationship to Logic Programming} \label{SecLP}

Autoepistemic logic is closely related to several semantics for logic
programs with negation. It is well-known that both stable and 
supported models of logic programs can be described as expansions
of appropriate translations of programs into modal theories (see, for
instance, \cite{mt93}). In this section, we briefly discuss connections
of the semantics defined by the least point of the operator $\DER{}$
to some 3-valued semantics of logic programs. The details will be
provided in a forthcoming work.

Given a logic programming clause 
$$r  = a\leftarrow b_1,\ldots,b_k,\n(c_1),\ldots,\n(c_m),$$
define:
\[
\aela{r} = b_1\wedge\ldots\wedge b_k\wedge\neg Kc_1\wedge\ldots\wedge\neg
Kc_m \supset a
\]
and 
\[
\aelb{r} = Kb_1\wedge\ldots\wedge Kb_k\wedge\neg Kc_1\wedge\ldots\wedge\neg
Kc_m \supset a
\]
Embeddings $\aela{}$ and $\aelb{}$ can be extended to logic programs $P$.

Let $B$ be a belief pair. Define the {\em projection}, $\Proj{B}$, 
as the 3-valued interpretation $I$ such that $I(p) = \HH{B}(K(p))$. 

It turns out that fixpoints of the operator $\DER{\aela{P}}$
($\DER{\aelb{P}}$, respectively) precisely
correspond to 3-valued stable (supported, respectively) models of 
$P$ (the projection function $\Proj{\cdot}$ establishes 
the correspondence). Moreover, complete fixpoints of $\DER{\aela{P}}$ 
describe 2-valued stable (supported, respectively) models of $P$.
Finally, the least fixpoint of $\DER{\aelb{P}}$ captures the
Fitting-Kunen 3-valued semantics of a program $P$, and the least 
fixpoint of $\DER{\aela{P}}$ captures the well-founded semantics of $P$.

\section*{Acknowledgments}
This work was partially supported by the NSF grants IRI-9400568 and
IRI-9619233
{\small
\bibliographystyle{aaai}
\bibliography{/a/al/u/d5/csfac/mirek/logic/nonmonlog}
}
\end{document}